\documentstyle[12pt]{article}
\input psfig

\newcommand{\be}{\begin{equation}}
\newcommand{\ee}{\end{equation}}
\newcommand{\bea}{\begin{eqnarray}}
\newcommand{\eea}{\end{eqnarray}}
\newcommand{\half}{{\scriptstyle{{1\over 2}}}}
\newcommand{\quart}{{\scriptstyle{{1\over 4}}}}

\newcommand{\Tr}{\mbox{\,Tr\,}}
\newcommand{\cF}{{\cal F}}
\newcommand{\cD}{{\cal D}}
\newcommand{\cO}{{\cal O}}
\newcommand{\RE}{{\rm Re~}}

\def\Journal#1#2#3#4{{#1} {#2} (#4) #3}
\def\NPB{{Nucl. Phys.} B}
\def\PLB{{Phys. Lett.} B}
\def\PRL{Phys. Rev. Lett.}
\def\CMP{Comm. Math. Phys.}
\def\ANP{Ann. Phys.}
\def\PRD{{Phys. Rev.} D}

\newcommand{\basispl}{
   \put(-.5,-.5){\line(1,0){1}}
   \put(.5,-.5){\line(0,1){1}}
   \put(.5,.5){\line(-1,0){1}}
   \put(-.5,.5){\line(0,-1){1}}}
\newcommand{\plaq}{\setlength{\unitlength}{.5cm}\raisebox{-.2cm}{
   \begin{picture}(1.2,1.2)(-.6,-.6)
   \basispl
   \put(-.5,-.5){\circle*{.2}}
   \put(-.5,.5){\circle*{.2}}
   \put(.5,-.5){\circle*{.2}}
   \put(.5,.5){\circle*{.2}}
   \put(.5,0){\vector(0,1){0}}
   \put(-.6,-.6){\makebox(0,0)[tr]{\footnotesize $x$}}
   \put(-.55,0){\makebox(0,0)[r]{\footnotesize $\nu$}}
   \put(0,-.55){\makebox(0,0)[t]{\footnotesize $\mu$}}
   \end{picture}}}
\newcommand{\twooneplaq}{\setlength{\unitlength}{.5cm}
   \raisebox{-.2cm}{
   \begin{picture}(2.2,1.2)(-1.1,-.6)
   \put(-1,-.5){\line(1,0){2}}
   \put(-1,.5){\line(1,0){2}}
   \put(-1,-.5){\line(0,1){1}}
   \put(1,-.5){\line(0,1){1}}
   \multiput(-1,-.5)(1,0){3}{\circle*{.2}}
   \multiput(-1,.5)(1,0){3}{\circle*{.2}}
   \put(-1.1,-.6){\makebox(0,0)[tr]{\footnotesize $x$}}
   \put(-1.05,0){\makebox(0,0)[r]{\footnotesize $\nu$}}
   \put(-.3,-.55){\makebox(0,0)[t]{\footnotesize $\mu$}}
   \put(1,0){\vector(0,1){0}}
   \end{picture}}}
\newcommand{\plaqa}{\setlength{\unitlength}{.5cm}\raisebox{-.2cm}{
   \begin{picture}(1.2,1.2)(-.6,-.6)
   \basispl
   \put(-.5,-.5){\circle*{.2}}
   \put(-.5,.5){\circle*{.2}}
   \put(.5,-.5){\circle*{.2}}
   \put(.5,.5){\circle*{.2}}
   \end{picture}}}
\newcommand{\hookplaq}{\setlength{\unitlength}{.5cm}
   \raisebox{-.3268cm}{
   \begin{picture}(1.7071,1.7071)(-.7071,-.7071)
   \put(0,0){\line(0,1){1}}
   \put(0,1){\line(1,0){1}}
   \put(1,1){\line(0,-1){1}}
   \put(-.7071,-.7071){\line(1,0){1}}
   \put(0,0){\line(-1,-1){.7071}}
   \put(1,0){\line(-1,-1){.7071}}
   \multiput(0,0)(1,0){2}{\circle*{.2}}
   \multiput(0,1)(1,0){2}{\circle*{.2}}
   \multiput(-.7071,-.7071)(1,0){2}{\circle*{.2}}
   \multiput(0,0)(.25,0){4}{\circle*{.03}}
   \put(1,0.5){\vector(0,1){0}}
   \put(-.8,-.8){\makebox(0,0)[tr]{\footnotesize $x$}}
   \put(-.2,.6){\makebox(0,0)[r]{\footnotesize $\nu$}}
   \put(-.2,-.8){\makebox(0,0)[t]{\footnotesize $\mu$}}
   \put(.9,-.4){\makebox(0,0)[t]{\footnotesize $\lambda$}}
   \end{picture}}}
\newcommand{\cornplaq}{\setlength{\unitlength}{.5cm}
   \raisebox{-.3268cm}{
   \begin{picture}(1.7071,1.7071)(-.7071,-.7071)
   \put(-.7071,-.7071){\line(0,1){1}}
   \put(0,1){\line(1,0){1}}
   \put(1,1){\line(0,-1){1}}
   \put(-.7071,-.7071){\line(1,0){1}}
   \put(0,1){\line(-1,-1){.7071}}
   \put(1,0){\line(-1,-1){.7071}}
   \put(-.7071,-.7071){\circle*{.1}}
   \put(-.7071,.2929){\circle*{.2}}
   \multiput(0,0)(1,0){2}{\circle*{.2}}
   \multiput(0,1)(1,0){2}{\circle*{.2}}
   \multiput(-.7071,-.7071)(1,0){2}{\circle*{.2}}
   \multiput(0,0)(.25,0){4}{\circle*{.03}}
   \multiput(0,0)(0,.25){4}{\circle*{.03}}
   \multiput(0,0)(-.1768,-.1768){4}{\circle*{.03}}
   \put(1,0.5){\vector(0,1){0}}
   \put(-.8,-.8){\makebox(0,0)[tr]{\footnotesize $x$}}
   \put(-.8,-.2){\makebox(0,0)[r]{\footnotesize $\nu$}}
   \put(-.2,-.8){\makebox(0,0)[t]{\footnotesize $\mu$}}
   \put(.9,-.4){\makebox(0,0)[t]{\footnotesize $\lambda$}}
   \end{picture}}}
\newcommand{\twoplaq}{\setlength{\unitlength}{1cm}\raisebox{-.5cm}{
   \begin{picture}(1.2,1.2)(-.6,-.6)
   \basispl
   \put(-.5,-.5){\circle*{.1}}
   \put(-.5,.5){\circle*{.1}}
   \put(.5,-.5){\circle*{.1}}
   \put(.5,.5){\circle*{.1}}
   \put(0,-.5){\circle*{.1}}
   \put(0,.5){\circle*{.1}}
   \put(.5,0){\circle*{.1}}
   \put(-.5,0){\circle*{.1}}
   \put(.5,-.2){\vector(0,1){0}}
   \put(-.55,-.55){\makebox(0,0)[tr]{\footnotesize $x$}}
   \put(-.55,-.2){\makebox(0,0)[r]{\footnotesize $\nu$}}
   \put(-.2,-.55){\makebox(0,0)[t]{\footnotesize $\mu$}}
   \end{picture}}}

\def\pho{\hphantom{1}}
\def\mystrut{{\vrule height 15pt depth 4pt width 0pt}}

\begin{document}
\vskip-1cm
\hfill INLO-PUB-12/96
\vskip5mm
\begin{center}
{\LARGE{\bf{\underline{A Monte Carlo study of old, new and}\\
\underline{tadpole improved actions}}}}\\
\vspace{1cm}
{\large Margarita Garc\'{\i}a P\'erez, Jeroen Snippe and Pierre van Baal 
} \\
\vspace{1cm}
Instituut-Lorentz for Theoretical Physics,\\
University of Leiden, PO Box 9506,\\
NL-2300 RA Leiden, The Netherlands.\\ 
\end{center}
\vspace*{5mm}{\narrower\narrower{\noindent
\underline{Abstract:} 
Scaling of mass ratios in intermediate volumes, obtained with improved
SU(2) lattice actions is tested against analytic results for the Wilson
and continuum action. A new improved action is introduced by adding a 
$2\times 2$ plaquette to the Symanzik action. Completing a square leads to 
a covariant propagator that simplifies perturbative calculations. Data is 
presented on lattices of size $4^3\times 128$, with lattice spacings of 
approximately 0.02 and 0.12 fermi. For the latter case no further improvement 
as compared to the tree-level action was observed when including the 
Lepage-Mackenzie tadpole correction to the one-loop improved L\"uscher-Weisz 
Symanzik action.}\par}
\section{Introduction}
Improvement of lattice actions aims at doing Monte Carlo simulations on 
coarser lattices, such that with a modest number of lattice spacings the 
physical volume is sufficiently large. But perhaps more importantly it should 
make extrapolations to the continuum limit more reliable, as has been one of 
the main objectives in the non-perturbative determination of the running 
coupling constant~\cite{alpha}. Here we consider the Symanzik improvement 
scheme~\cite{sym}, which is designed to remove lattice artefacts by adding 
irrelevant operators to the lattice action, whose coefficients are tuned by 
requiring spectral quantities to be improved to the relevant order (on-shell 
improvement~\cite{luwe}).  For Symanzik improvement to work it seemed 
that unreasonably small values of the bare coupling constant were required. 

Mean field inspired Symanzik improvement~\cite{lema} was introduced
to beat the bad convergence of perturbative expansions in the bare coupling
constant. In particular the Parisi mean field coupling~\cite{par} defined in
terms of the plaquette expectation value is seen to improve considerably the
approach to asymptotic scaling. Despite some attempts~\cite{per} no good
theoretical understanding for this is available. The tadpole prescription also
includes corrections~\cite{lema} to the coefficients in the Symanzik improved 
action, which can be seen as a mean field renormalization of the link 
variables on the lattice. Only phenomenological arguments have been provided 
to support this. Standard tests in pure gauge theories involve restoration of 
rotational invariance~\cite{lema}. More involved, but of direct physical 
relevance, are the tests in charmonium spectroscopy~\cite{corn}, used to 
extract a value of the strong coupling constant~\cite{char}. 

We here stress the necessity of improving scaling, rather than asymptotic 
scaling, which in spectroscopy is less important since one has to set the scale 
by one of the masses or the string tension. In this sense our study is 
complementary to that of ref.~\cite{morn}.
Although one is ultimately interested in the infinite volume limit, 
from the point of studying the approach to the continuum limit a finite
volume provides a useful tool. If improvement fails there, it sheds doubt on
results in large volumes (when successful, however, one does not imply the 
other). Perhaps a somewhat inappropriate comparison is that we consider our 
study as a well controlled laboratory experiment, where conditions are 
manipulated so as to rule out as much as possible external disturbances.

The setup of this letter is to first introduce and motivate the new improved
action. It simplifies certain perturbative calculations, and provides in part 
an analytic test of improvement in a small volume for which we present the
effective potential in a constant abelian background field. Also the Lambda
parameter of the new improved action is related to that of the Wilson action.
We then present our Monte Carlo data at very small and intermediate volumes 
and end with conclusions.  Details of the analytic study will be presented 
elsewhere (preliminary material and some further discussion can also be found 
in two communications to conferences~\cite{us}).

\section{Square Symanzik action}
There is a large redundancy in choosing an improved action, when parametrized 
in terms of Wilson loops. We shall use this to allow for a simplified 
``covariant'' gauge choice, achieved by adding to the L\"uscher-Weisz (LW)
Symanzik action a $2\times2$ plaquette.
\bea
S(\{c_i\})&\equiv&\sum_x\Tr\Biggl\{c_0\sum_{\mu\neq\nu}\left\langle1-\plaq\ 
\right\rangle+2c_1\sum_{\mu\neq\nu}\left\langle1-\twooneplaq\ \right\rangle+
\frac{4}{3}c_2\sum_{\mu\neq\nu\neq\lambda}\left\langle1-\cornplaq 
\ \right\rangle\nonumber\\&&\hskip3cm+4c_3\sum_{\mu\neq\nu\neq\lambda}
\left\langle1-\hskip-3mm\hookplaq\ \right\rangle
+c_4\sum_{\mu\neq\nu}\left\langle1-\twoplaq \ \right\rangle\Biggr\}.
\eea
The $<>$ imply averaging over the two opposite directions for each of the 
links, called ``clover'' averaging in ref.~\cite{mgp}. Numerical factors were 
chosen to agree with earlier conventions~\cite{luwe}. Note that sometimes 
$c_2$ and $c_3$ are interchanged in the literature~\cite{mgp,wei}. Here 
$c_4$ is assigned to the $2\times2$ plaquette. 

The number of parameters required to improve the action to a certain order is 
simply determined from the number of gauge and hypercubic invariant operators 
that one can write down up to that order (read off from the dimension of the 
operator). For pure gauge theories
there is only one operator of dimension zero and three of dimension two.
One of these is redundant as it can be removed by a field 
redefinition, which can also be implemented at the level of the Wilson loop 
representation. It allows one to choose~\cite{luwe} $c_3=0$. 

As usual we relate lattice and continuum fields by 
$U_\mu(x)=P\exp(\int_0^a A_\mu(x+s\hat\mu)ds)$. This gives the following
expansion for the lattice action~\cite{mgp,wei}
\bea
S(\{c_i\})
&=&-\frac{a^4}{2}(c_0+8c_1+8c_2+16c_3+16c_4)\sum_{x,\mu,\nu}
\Tr(F_{\mu\nu}^2(x))\nonumber\\&&+a^6(\frac{c_2}{3}\!+\!c_3)\sum_{x,\mu,
\nu,\lambda}\Tr(\cD_{\mu}F_{\mu\lambda}(x)\cD_{\nu}F_{\nu\lambda}(x))
+a^6\frac{c_2}{3}\sum_{x,\mu,\nu,\lambda}\Tr((\cD_\mu F_{\nu\lambda}(x))^2)
\nonumber\\&&+\frac{a^6}{12}(c_0+20c_1-4c_2+4c_3+64c_4)\sum_{x,\mu,\nu}
\Tr(\cD_{\mu}F_{\mu\nu}(x))^2+\cO(a^8)\ .
\label{eq:weisac}
\eea
To fix the definition of the coupling constant one imposes $(c_0+8c_1+8c_2+
16c_3+16c_4)=1$. Computing two particular spectral quantities as a function 
of these parameters allows one to determine these coefficients. At tree-level 
the conventional choice amounts to putting $c_0=5/3$, $c_1=-1/12$ and 
$c_2=c_3=c_4=0$. The one-loop ($\cO(g_0^2)$) correction to these coefficients 
was computed by L\"uscher and Weisz~\cite{luwe}. For $c_4\neq0$ a similar 
calculation is in the process of being completed by one of us. At tree-level 
we have fixed $c_4$ by the following requirement. When expanding the action to 
quadratic order in the lattice field $q_\mu(x)$, defined by 
$U_\mu(x)=\exp(q_\mu(x))$, one finds
\bea
S_2&=&\sum_{x,\mu,\nu}-\half\Tr[c_0(\partial_\mu q_\nu(x)-\partial_\nu
q_\mu(x))^2+2c_1\{(2+\partial_\mu)(\partial_\mu q_\nu(x)-\partial_\nu q_\mu(
x))\}^2\nonumber\\&&\hskip2cm+c_4\{(2+\partial_\nu)(2+\partial_\mu)(
\partial_\mu q_\nu(x)-\partial_\nu q_\mu(x))\}^2]\ ,
\eea
where $\partial_\mu$ is the lattice difference operator $\partial_\mu\varphi(x)
\equiv\varphi(x+\hat\mu)-\varphi(x)$. If we now choose 
\be 
c_4\equiv z^2 c_0,\quad z\equiv c_1/c_0,
\ee 
we can complete squares and obtain a simple gauge fixing function
\be
\cF_{gf}(x)\equiv\sqrt{c_0}\sum_{\mu}\partial_\mu^\dagger
\left(1+z(2+\partial_\mu^\dagger)(2+\partial_\mu)\right)q_\mu(x).
\ee
It is for this reason we propose to call the new improved action the square 
Symanzik action.
At tree-level one finds
\be
c_0=16/9,\quad c_1=-1/9,\quad c_2=0,\quad c_3=0,\quad c_4=1/144.
\ee
An amusing, and potentially useful feature is that the relation $c_4c_0=c_1^2$
is not affected by tadpole corrections, where one replaces 
$U_\mu(x)$ by $U_\mu(x)/u_0$, with $u_0$ the fourth root of the average value 
of the plaquette.
\be 
u_0^4=\RE k^{-1}\Tr\left\langle\plaqa~\right\rangle.
\ee
Here $k$ is the number of colors. For values of $u_0\neq 1$ one easily
finds $z=-1/(16u_0^2)$ and $c_0=1/(1+4z)^2$. 

In the covariant gauge the propagators for the ghost and vector fields
are simply given by
\bea
{\rm Ghost}:&&P(k)=\frac{1}{\sqrt{c_0}\sum_\lambda\left(4\sin^2(k_\lambda/2)
+4z\sin^2k_\lambda\right)}\ ,\nonumber\\{\rm Vector}:&&P_{\mu\nu}(k)=
\frac{P(k)\delta_{\mu\nu}}{\sqrt{c_0}\left(1+4z\cos^2(k_\mu/2)\right)}\ .
\eea
It illustrates an important feature of improved actions with more than nearest
neighbor couplings in the time direction: unphysical poles appear at the scale 
of the cutoff, $1/a$. For low energy physics they are harmless~\cite{ham} in 
the same way Pauli-Villars regulator fields are harmless at energies below the 
scale of the cutoff. However, on the lattice these spurious poles are more 
cumbersome to handle as they do not simply appear in loops (i.e. vertices do 
not preserve something like ghost number). Nevertheless, there is a way of 
separating off their contributions~\cite{us}. Each of the propagators can be 
factorized in the sum of two or three standard (single pole) lattice 
propagators, $P_s\equiv1/(4\sin^2(\half k_0)+\omega_s^2(\vec k))$,
\bea
P(k)&=&Z(\vec k)(P_-(k)-P_+(k)),\\ P_{\mu\nu}(k)&=&\delta_{\mu\nu}(
Z^-_\mu(\vec k)P_-(k)-Z^+_\mu(\vec k)P_+(k)+Z^0_\mu(\vec k)P_0(k)),\nonumber
\eea
It is straightforward to derive the explicit expressions for the $Z$ factors and energies $\omega$ from eq.~(8). Note that $Z_j^0=0$ and that (for $u_0=1$)
$\omega_+^2(\vec 0)=\omega_-^2(\vec 0)=12$. The spurious poles in this case
occur at an energy $2{\rm asinh}(\sqrt{3})/a$. 

One particularly simple test of improvement is achieved by computing for
SU(2) the effective potential for a static abelian zero-momentum background 
field, $\hat U_j=\exp(\half iC_j\sigma_3/N)$ and $\hat U_0=1$, that is a 
solution of the (lattice) equations of motion. Introducing the quantum 
fluctuations through $U_\mu(x)=e^{\hat q_\mu(x)}\hat U_\mu$ one easily 
diagonalizes the quadratic fluctuation operator in the covariant gauge 
\be
\hat\cF_{gf}\equiv\sqrt{c_0}\sum_{\mu}\hat D_\mu^\dagger\left(1+z(2
+\hat D_\mu^\dagger)(2+\hat D_\mu)\right)\hat q_\mu(x),
\ee
where $\hat D_\mu\varphi(x)\equiv \hat U_\mu\varphi(x+\hat\mu)\hat 
U^\dagger_\mu-\varphi(x)$. Due to the background field, momenta will be shifted
to $\vec k^s=(2\pi\vec n+s\vec C)/N$, where $s=0$ for the isospin
neutral and $s=\pm1$ for the isospin charged components of quantum fields.
The eigenvalues can be directly read off from eq.~(8) and one finds 
\be
V^{\rm ab}_1(\vec C)=N\sum_{\vec n\in Z_N^3}\Biggl\{\sum_i\log\left(\lambda_i
\right)+4{\rm asinh}\left(2u_0\sqrt{1+4z+\frac{\omega^2}{2}+
\omega\sqrt{1+\frac{\omega^2}{4}}}\ \right)\Biggr\},
\ee
with $\lambda_j=1+4z\cos^2(\half k_j^+)$ and $\omega^2=\sum_j4\lambda_j
\sin^2(\half k^+_j)$. The result, normalized to $V_1^{\rm ab}(\vec 0)\!=\!0$, 
is plotted in figure 1 for $u_0=1$ (together with the effective potential for 
the Wilson action, obtained by taking $z=0$). At $N=6$ we can not distinguish 
the result from the continuum at the scale of this figure. 

We can use the abelian background field also to compute the one-loop correction
to the tree-level kinetic term $\half g_0^{-2}(dC_i(t)/dt)^2$, which yields 
$\half(g^{-2}+\alpha_1)(dC_i(t)/dt)^2$. In the continuum limit 
$g^{-2}=g_0^{-2}-11\log(N)/12\pi^2$ is kept fixed while sending the number of 
lattice spacings, $N$, to infinity. An analytic expression for $\alpha_1(N)$ 
was found in terms of a sum over spatial momenta, which reduces to the result 
for the Wilson action~\cite{vb} at $z=0$. Computing this sum for one hundred 
lattices and fitting the result to a polynomial in $1/N$ we find $\alpha_1=
-0.0340012235(1)$ for $z=-1/16$ and $\alpha_1=-0.1648688946(1)$ for $z=0$.
From the difference one determines the ratios of the Lambda parameters
between the square Symanzik ($\Lambda_{S^2}$) and the Wilson ($\Lambda_W$)
actions. One can also compute the one-loop correction for 
$\quart g_0^{-2}\Tr F_{ij}^2$, which gives an identical result for the 
Lambda ratios. Alternatively, we used the heavy quark potential 
method~\cite{wei}, which also allowed us to extract the Lambda ratios 
for SU(3). We quote the following result
\be
\Lambda_{S^2}/\Lambda_W[{\rm SU(2)}]=4.0919901(1),\quad 
\Lambda_{S^2}/\Lambda_W[{\rm SU(3)}]=5.2089503(1).
\ee
In addition, the one-loop perturbative expansions of the SU($k$) expectation 
values for an $a\times b$ plaquette, $U(P(a,b))$, are given by
\be
\left\langle\RE k^{-1}\Tr U(P(a,b))\right\rangle=1-\quart 
g_0^2(k-k^{-1})w(a,b).
\ee
For the square Symanzik action (eq.~(6)) we find
\be
w(1,1)=0.3587838551(1),\quad w(1,2)=0.6542934512(1),\quad 
w(2,2)=1.0887235337(1).
\ee

\section{Monte Carlo data}
We wish to determine in small volumes the mass for the scalar ($A_1^+$) and
tensor glueballs, the latter split due to the breaking of rotational invariance
in the doublet $E^+$ and the triplet $T_2^+$. Also the energies of the electric
flux (``torelon'') states with one, two and three units of electric flux 
($e_i$, $i=1,2,3$) will be measured. In addition we consider the states 
with two ($T_{11}^+$ or $B(110)$) and three ($T_2^+(111)$) units of electric 
flux that have $T^+_2$ quantum numbers (negative parity in two directions of 
electric flux, symmetrized in those two directions). See ref.~\cite{mic} for
details and further references.

The size of the lattice used is $4^3\times 128$ and masses $m$ are converted
to dimensionless parameters into $z=mL$; in lattice units we hence multiply 
the mass with the number of lattice sites in the spatial directions.
In large volumes one should have $z_{e_k}=\sigma L^2\sqrt{k}$, where $\sigma$ 
is the infinite volume string tension. This is why we will consider the 
rations $\sqrt{z_{e_k}}/z_{A_1^+}$. These and other mass ratios will be 
plotted as a function of $z_{A_1^+}$. The analytic result~\cite{vb} derived 
by diagonalizing an effective Hamiltonian to describe low-lying states is valid 
up to $z_{A_1^+}\sim 5$, after which degrees of freedom that were integrated 
out perturbatively will receive non-perturbative contributions~\cite{mgp}. The 
breakdown will occur at smaller volumes for higher excited states. 

For the Wilson action we have chosen $\beta=3.0$ and $\beta=2.4$; for the 
improved actions $\beta$ was tuned to yield results in roughly the same 
physical volume. These parameters correspond to lattice spacings of 
approximately $a=0.018$ and  $a=0.12$ fermi. For the smallest of these two, 
one expects tree-level improvement to be effective and we have therefore not 
tadpole corrected the actions in this case. Note that for these small volumes 
one finds from the analytic results that the lattice artefacts in the mass 
ratios are quite much bigger~\cite{vb} than in larger volumes. Data was taken 
for both the LW and square Symanzik actions, and as a test on our programs
also for the Wilson action for which we can compare with available high 
precision data~\cite{mic}. 

At the larger volume we concentrated our attention to the LW Symanzik action 
with tree-level and tadpole corrected one-loop values of the coefficients. 
We verified that there is no observable volume dependence of $u_0$ by 
comparing its value with the one on an $8^3\times64$ lattice (the difference 
was less than 0.3\%, consistent with zero within statistical errors). 
Following the prescription of refs.~\cite{lema,corn,morn} we took for SU(2)
\bea
&&c_0=5/3,\quad c_1=-(1+0.2227\alpha_s(u_0))/(12u_0^2),\quad c_2=-0.02224\times
5\alpha_s/(3u_0^2),\quad c_4=0,\nonumber\\
&&\hskip4cm\alpha_s(u_0)\equiv-(4\log u_0)/1.725969,
\eea
obtained from the one-loop coefficients determined by L\"uscher, Weisz and 
Wohlert~\cite{luwe,wei}. Substituting these coefficients in eq.~(1) and
multiplying by $\beta/4\equiv1/g_0^2$ gives the action we used for
our simulations. We do not absorb the tree-level value of $c_0$ in the 
definition of the coupling constant, as was done in ref.~\cite{corn,morn}. 
When using the convention of eq.~(15) the standard two-loop relation between 
$\beta$ and $a\Lambda$ needs no modification. But the Lambda parameter has to 
be corrected for the fact that the L\"uscher-Weisz choice of coupling amounts 
to multiplying eq.~(1) by $(g_0^{-2}+0.08112)$, so as to compensate for the 
one-loop correction to $c_0$.

In the intermediate volumes we have used for our simulations, masses remain 
small compared to the spurious unphysical poles. This allows us to use the 
variational approach~\cite{var} to increase the overlap of the the states to 
be measured. We have been able to extract clean signals. On very coarse 
lattices where masses would no longer be small in lattice units, one looses 
the signal in the noise too early to extract it reliably, whereas also the 
variational method is no longer well founded. Recently these problems were 
tackled by using anisotropic lattices~\cite{morn}, well known from finite 
temperature studies~\cite{kar}. Only implementing improvement for the spatial 
directions will in addition remove the problems with a non-hermitian transfer 
matrix. 

The raw data are listed in tables 1 and 2, based on performing the variational 
analysis on the second time slice. We have verified that the result is stable 
against performing the variational analysis on the first time slice. We used 
3 to 8 operators, as defined in ref.~\cite{mic}, for the variational 
analysis. They were computed in terms of Teper-fuzzed links~\cite{tep}. Only 
for the determination of the scalar glueball mass at $a\sim 0.02$ fermi the
variational analysis was important, in most other cases a single {\em but 
Teper-fuzzed} operator was sufficient to obtain accurate results. 

Another issue is that for $a\sim 0.02$ fermi the small value of the coupling 
gives rise to large autocorrelations that can affect the energies of electric 
flux. In most cases we found it useful to correct for this by eliminating data 
for which the average of the spatial Polyakov loops over the 128 time slices 
(and a few heat bath updates) was bigger in absolute value than one half. Our 
results for the Wilson case at $\beta=3$ agree to high accuracy with those 
reported by Michael~\cite{mic}.

Because of the availability of analytic results, it is not necessary to 
exactly tune the different actions to the same physical volume. Nevertheless 
in particular for the data at $a=0.12$ fermi we made an effort to tune 
parameters appropriately, as we can make a stronger point when directly 
comparing lattice data at the same physical volume. The value of $u_0$ is 
determined self consistently~\cite{corn,morn}, adjusting with the help of the 
Ferrenberg-Swendsen trick~\cite{fs} the input value of $u_0$ to agree with its
measured value. This only requires little Monte Carlo time. The results of 
tables 1 and 2 are presented in figure 2 to compare with the analytic results 
for the continuum (solid curves) and for the Wilson action on a lattice
of size $4^3\times\infty$. We have used approximately 160 hours of CPU time on 
a Cray C98 to generate and analyze the data presented in this paper. 
Computational overheads for improved actions amount to a factor 3 for the 
LW and 4 for the square Symanzik action over the standard Wilson action.

\section{Discussion}
As was to be expected, at lattice spacings of approximately $0.02$ fermi
($z_{A_1^+}\sim 2$), tree-level improvement is seen to bring the lattice 
results quite close to those of the continuum, both for the LW and square 
Symanzik actions. In both cases the improvement is considerable.

Also at lattice spacings around $0.12$ fermi and volumes of approximately 
$0.48$ fermi ($z_{A_1^+}\sim 4$), the agreement of the Wilson action lattice 
data with the corresponding analytic results is in general very good for the 
lowest lying states. The difference in the analytic result between the 
continuum and Wilson lattice action gives an indication how far the improved 
data is removed from the continuum result. Significant improvement is observed 
in some of the cases, in particular for $z_{T^+_{11}}/z_{A_1^+}$, approaching 
the continuum analytic result.

The most salient feature of our data is that tadpole correction has no 
significant effect on the tree-level improved data for the ratios. 
Perhaps for the cases where tree-level improvement is already significant
this is what one would want, but our results show some instances where 
tree-level improvement has no effect {\em and} the tadpole correction is 
of no help either.

In particular we note that the ratio $\sqrt{z_{e_1}}/z_{A_1^+}$, measured to 
an accuracy of better than 1.5\%, deviates from its continuum value by 5-6\%. 
For this quantity tree-level improvement {\em as well as tadpole corrected} 
one-loop improvement are unable to show deviations from the Wilson result. 
This result puts some doubt on the usefulness of the tadpole correction 
for careful extrapolations of mass ratios to the continuum limit. 

One might object that the lattice spacing we have used to implement the 
tadpole correction, $a=0.12$ fermi, is not really large enough. We have 
certainly not probed lattice spacings as large as $a=0.4$ fermi, that have 
been advertised~\cite{corn}. Nevertheless for $a=0.12$ fermi, $u^4_0= 
0.6819(1)$ and significantly deviates from $1$. The correction to $c_1$ at 
these parameters is 27\% with respect to its tree-level value (without 
tadpole correction it would have been 17\%). 

\section*{Acknowledgements}

This work was supported in part by FOM and by a grant from NCF for use 
of the Cray C98. We are grateful to Colin Morningstar and Mark Alford
for discussions and correspondence on implementing tadpole improvement and
to Mike Teper for correspondence on measuring masses.

\newpage
\hbox{\hskip3.4cm  Wilson action, $\beta=3.0$ \hskip3cm }
\vskip.5mm
\hbox to \hsize{\hfil\vbox{\offinterlineskip
\halign{&\vrule#&\ $#\mystrut$\hfil\ \cr
\noalign{\hrule}
&{\rm Rep.} && \# {\rm op.} &&1/0&&2/1&&3/2&&4/3&\cr
height 4pt&\omit&&\omit&&\omit&&\omit&&\omit&&\omit&\cr
\noalign{\hrule}
height 4pt&\omit&&\omit&&\omit&&\omit&&\omit&&\omit&\cr
&A_1^+ && 5 &&2.105(9)  && {\bf 2.03(1)}&& 2.02(2)&& 2.08(3) &\cr
&E^+   && 3 &&1.743(5)  && {\bf 1.703(9)}&& 1.71(1)&& 1.70(2) &\cr
&T_2^+ && 3 &&3.315(9)  && 3.25(2)&& {\bf 3.21(4)}&& 3.2(1) &\cr 
&e_1   && 3 &&0.277(3)  && {\bf 0.269(5)}&& 0.269(6)&& 0.270(7) &\cr 
&e_2   && 3 &&0.588(5)  && {\bf 0.575(7)}&& 0.576(9)&& 0.58(1) &\cr  
&e_3  && 3 &&0.978(5)  && {\bf 0.962(8)}&& 0.97(1)&& 0.97(2) &\cr
\noalign{\hrule}}}\hfil}
\vskip.5mm
\hbox{\hskip3.4cm  LW Symanzik action, $\beta=2.374$
 \hskip3cm } 
\vskip.5mm
\hbox to \hsize{\hfil\vbox{\offinterlineskip
\halign{&\vrule#&\ $#\mystrut$\hfil\ \cr
\noalign{\hrule}
&{\rm Rep.} && \# {\rm op.} &&1/0&&2/1&&3/2&&4/3&\cr
height 4pt&\omit&&\omit&&\omit&&\omit&&\omit&&\omit&\cr
\noalign{\hrule}
height 4pt&\omit&&\omit&&\omit&&\omit&&\omit&&\omit&\cr
& A_1^+  && 6 &&1.89(1)  && 2.01(2)&& {\bf 2.03(3)}&& 2.09(5) &\cr
& E^+    && 3 &&1.626(6)  && {\bf 1.77(1)}&& 1.78(2)&& 1.80(3) &\cr 
& T_2^+  && 3 &&3.134(8)  && 3.43(2)&& {\bf 3.47(5)}&& 3.4(1) &\cr  
& e_1  && 3 &&0.307(3)  && 0.334(5)&& {\bf 0.339(6)}&& 0.342(8) &\cr  
& e_2  && 3 &&0.656(5)  && 0.718(8)&& {\bf 0.73(1)}&& 0.74(2) &\cr   
& e_3  && 3 &&1.104(6)  && 1.21(1)&& {\bf 1.24(1)}&& 1.26(2) &\cr
\noalign{\hrule}}}\hfil}
\vskip.5mm
\hbox{\hskip3.4cm  Square Symanzik action, $\beta=2.2013$
 \hskip3cm }
\vskip.5mm
\hbox to \hsize{\hfil\vbox{\offinterlineskip
\halign{&\vrule#&\ $#\mystrut$\hfil\ \cr
\noalign{\hrule} 
&{\rm Rep.} && \# {\rm op.} &&1/0&&2/1&&3/2&&4/3&\cr 
height 4pt&\omit&&\omit&&\omit&&\omit&&\omit&&\omit&\cr 
\noalign{\hrule} 
height 4pt&\omit&&\omit&&\omit&&\omit&&\omit&&\omit&\cr 
& A_1^+ && 6 &&2.15(1)  && {\bf 2.31(2)}&& 2.30(3)&& 2.29(5) &\cr 
& E^+  && 3 &&1.844(6)  && 2.00(1)&& {\bf 2.02(2)}&& 2.04(3) &\cr  
& T_2^+ && 3 &&3.56(1)  && {\bf 3.87(3)}&& 3.88(7)&& 3.8(3) &\cr   
& e_1 && 3 &&0.375(4)  && 0.405(6)&& {\bf 0.408(7)}&& 0.412(9) &\cr
& e_2 && 3 &&0.808(6)  && 0.877(9)&& {\bf 0.89(1)}&& 0.90(2) &\cr
& e_3 && 3 &&1.368(8)  && 1.49(1)&& {\bf 1.52(2)}&& 1.54(9) &\cr
\noalign{\hrule}}}\hfil}
\vskip5mm
{{\noindent 
Table 1: Values of $z=mL$ at a lattice spacing of approximately $0.02$ fermi 
for SU(2) on a $4^3\times 128$ lattice. We have performed  16000 measurements 
(25 heat-bath sweeps apart) for Wilson and 20000 for both LW and square 
Symanzik actions (10 sweeps apart). The entries in the table correspond to  
the representations of the cubic group, the number of operators used in the 
variational analysis and the effective masses extracted from $n/\ell$ ratios 
of correlation functions, i.e. $-\log\left(\sum_t<\cO(t+n)\cO(t)>/\sum_t
<\cO(t+\ell)\cO(t)>\right)$. Entries in boldface are taken as final estimates 
for figure 2. Errors have been analyzed using the jackknife method. 
}\par} 
\newpage
\hbox{\hskip3.2cm  LW Symanzik action, $\beta=1.83$       
 \hskip3cm } 
\vskip.5mm
\hbox to \hsize{\hfil\vbox{\offinterlineskip 
\halign{&\vrule#&\ $#\mystrut$\hfil\ \cr
\noalign{\hrule}
&{\rm Rep.} && \# {\rm op.} &&1/0&&2/1&&3/2&&4/3&\cr
height 4pt&\omit&&\omit&&\omit&&\omit&&\omit&&\omit&\cr
\noalign{\hrule}
height 4pt&\omit&&\omit&&\omit&&\omit&&\omit&&\omit&\cr
& A_1^+   && 7 &&3.71(1\ )  && 3.74(2)&& {\bf 3.78(5)}&& 3.9(2) &\cr
& E^+  && 7 &&3.212(9)  && {\bf 3.29(2)}&&  3.30(4)&& 3.3(1) &\cr
& T_2^+&& 3 &&6.13(1)  && {\bf 6.31(6)}&&  6.3(3) && 6.2(1.0)   &\cr
& e_1 && 7 &&0.813(6)\pho  && {\bf 0.84(1)}&& 0.84(1)&& 0.84(2) &\cr
& e_2 && 7 &&1.75(1)  && {\bf 1.80(2)}&& 1.80(3)&& 1.80(3) &\cr
& e_3 && 8 &&2.89(2)  && 3.05(3)&& {\bf 3.09(5)}&& 3.2(1) &\cr
& T_{11}^+&&7 &&1.857(6)   && {\bf 1.92(1)}&& 1.92(2)&& 1.91(3) &\cr
& T_2^+(111) && 7 &&2.67(1)  && {\bf 2.67(2)}&&  2.66(3)&& 2.67(7) &\cr
\noalign{\hrule}}}\hfil}
\vskip.5mm
\hbox{\hskip3.2cm  Tadpole corrected LW Symanzik action, $\beta=2.04$
 \hskip3cm } 
\vskip.5mm
\hbox to \hsize{\hfil\vbox{\offinterlineskip 
\halign{&\vrule#&\ $#\mystrut$\hfil\ \cr
\noalign{\hrule}
&{\rm Rep.} && \# {\rm op.} &&1/0&&2/1&&3/2&&4/3&\cr
height 4pt&\omit&&\omit&&\omit&&\omit&&\omit&&\omit&\cr
\noalign{\hrule}
height 4pt&\omit&&\omit&&\omit&&\omit&&\omit&&\omit&\cr
& A_1^+ && 7 &&4.07(1)  && {\bf 4.06(3)}&& 4.05(8)&& 3.8(2) &\cr
& E^+  && 7 &&3.366(6)  && {\bf 3.57(2)}&& 3.57(5)&& 3.6(1) &\cr
& T_2^+ && 3 &&6.28(2)   && {\bf 6.76(7)}&&  6.6(4)&& 6.5(1.0) &\cr
& e_1 && 7 &&0.889(5)  && {\bf 0.94(1)}&& 0.94(1)&& 0.94(2) &\cr
& e_2 && 7 &&1.920(1)  && {\bf 2.02(2)}&& 2.01(3)&& 1.98(4) &\cr
& e_3 && 7 &&3.402(14)  && 3.41(3)&& {\bf 3.38(4)}&& 3.21(7) &\cr
& T_{11}^+&&7 &&1.893(6)   && {\bf 2.06(1)}&& 2.07(3)&& 2.07(4) &\cr
& T_2^+(111) && 7 &&2.79(1)  && {\bf 2.87(2)}&&  2.86(4)&& 2.83(8) &\cr
\noalign{\hrule}}}\hfil}
\vskip5mm
{{\noindent
Table 2: The same as in  table 1 but for a lattice spacing of approximately 
$0.12$ fermi.  We have performed  40000  and 48000 measurements respectively 
for tree-level LW and tadpole corrected one-loop improved LW Symanzik actions, 
in both cases separated 2 heat-bath sweeps apart.}\par}
\newpage
\begin{figure}[htb]
\vspace{14cm}
\includegraphics{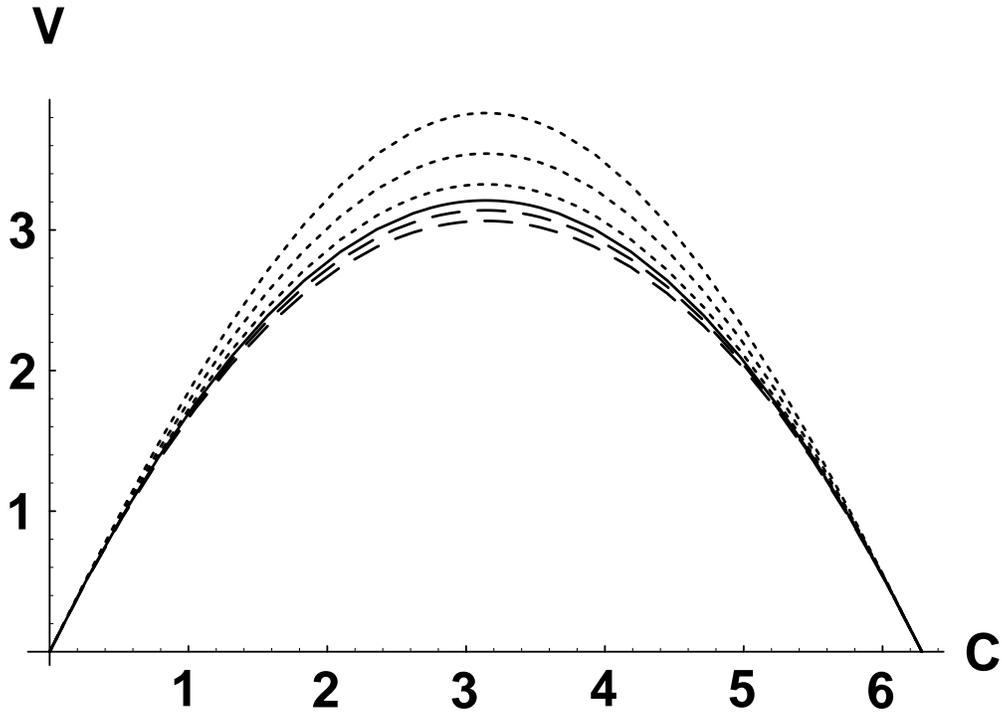}
\caption{
The SU(2) effective potential for a constant abelian background field
$\vec C=(C,0,0)$. The full line represents the continuum result (obtained by 
taking the number of lattice spacings $N\rightarrow\infty$). The lower two 
dashed curves are for the square Symanzik action ($u_0=1$) with $N=3$ and 4. 
The upper three dotted curves are for the Wilson action with $N=3,4$ and 6.}
\end{figure}
\begin{figure}[htb]
\vspace{15cm}
\includegraphics{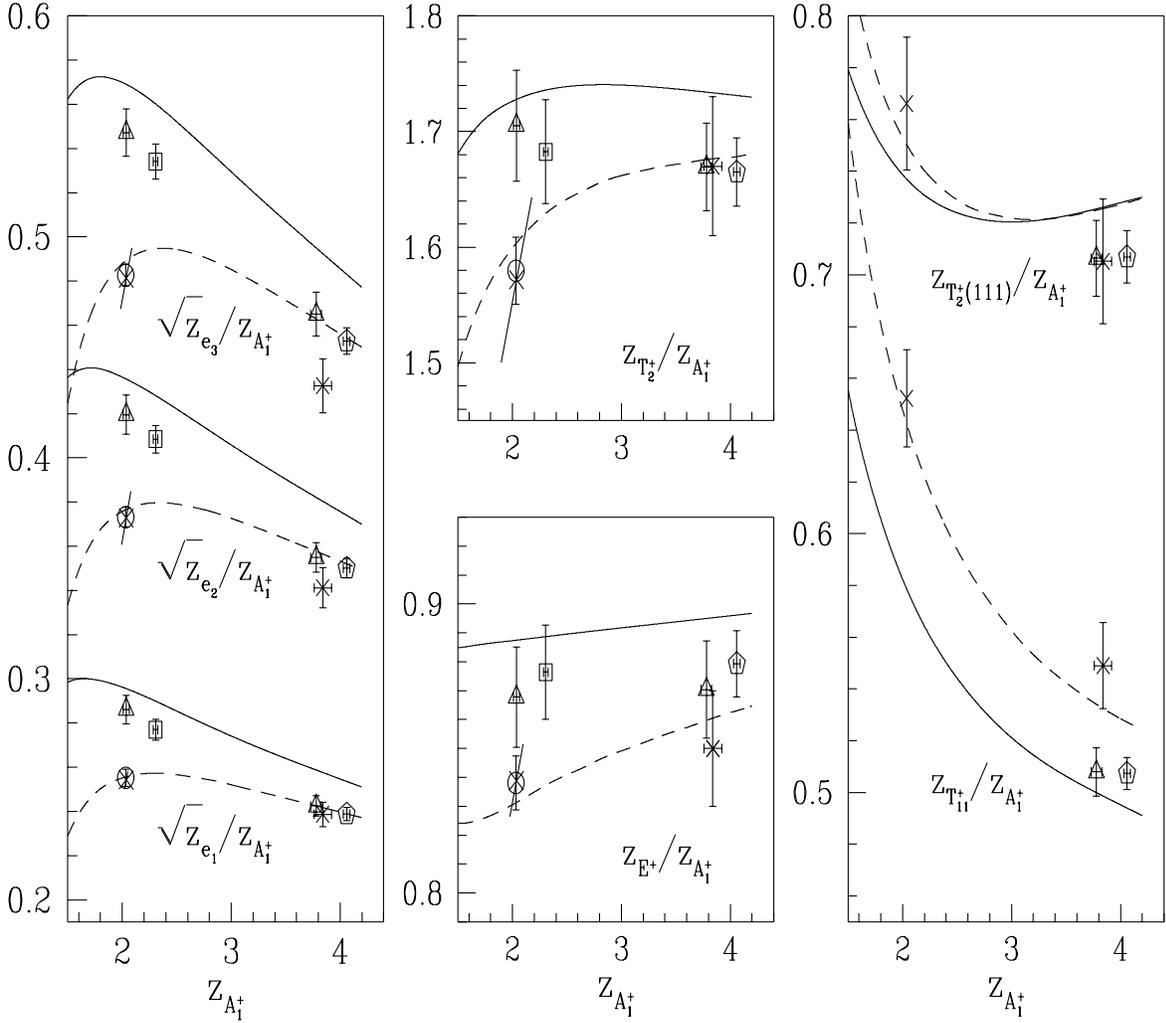}
\caption{SU(2) Monte Carlo data (see tables) on a $4^3\times 128$ lattice for 
the Wilson action (circles for our data and crosses for data by Michael~[15], 
with tilted error bars when data overlap), the LW Symanzik improved action 
(triangles), the square Symanzik action (squares) and the tadpole corrected 
one-loop LW Symanzik action (pentagons) at lattice spacings of approximately 
0.02 and 0.12 fermi. A comparison is made with analytic results for the 
continuum (solid lines) and Wilson action on a lattice of size 
$4^3\times\infty$ (dashed lines).}
\end{figure}

\newpage
\begin{thebibliography}{9}
\bibitem{alpha} M. L\"uscher, e.a., \Journal{\NPB}{413}{481}{1994}; 
G. de Divitiis, e.a., \Journal{\NPB}{437}{447}{1995}.
\bibitem{sym}K. Symanzik, \Journal{\NPB}{226}{187, 205}{1983}.
\bibitem{luwe}M. L\"{u}scher and P. Weisz, \Journal{\PLB}{158}{250}{1985};
\Journal{\NPB}{266}{309}{1986}; \Journal{\CMP}{97}{59}{1985}.
\bibitem{lema}G.P. Lepage and P.B. Mackenzie, \Journal{\PRD}{48}{2250}{1993}.
\bibitem{par}G. Parisi, in {\em High Energy Physics-1980}, eds. L. Durand and
L.G. Pondrom (American Institute of Physics, New York, 1981).
\bibitem{per}V. Periwal, \Journal{\PRD}{53}{2605}{1996}.
\bibitem{corn} G.P. Lepage, \Journal{\NPB}{(Proc.Suppl)47}{3}{1996};
M. Alford, e.a., \Journal{\PLB}{361}{87}{1995}.
\bibitem{char} A.X. El-Khadra, e.a., \Journal{\PRL}{69}{729}{1992}.
\bibitem{morn} C. Morningstar and M. Peardon, 
\Journal{\NPB}{(Proc.Suppl.)47}{258}{1996}; hep-lat/\break9606008, to appear 
in the proceedings of PANIC'96, Williamsburg, VA, 22-28 May, 1996.
\bibitem{us} M. Garc\'{\i}a P\'erez, J. Snippe and P. van Baal, 
hep-lat/9607007 and hep-lat/9608015, to appear in the proceedings of
the second workshop ``Continuous advances in QCD'', Univ. of Minnesota,
28-31 March, 1996 and in the proceedings of Lattice'96, Washington Univ.,
St. Louis, 4-8 July, 1996.
\bibitem{mgp}M. Garc\'{\i}a P\'erez, e.a., \Journal{\NPB}{413}{535}{1994}.
\bibitem{wei}P. Weisz, \Journal{\NPB}{212}{1}{1983}; P. Weisz and R. Wohlert,
\Journal{\NPB}{236}{397}{1984}.
\bibitem{ham}M. L\"{u}scher and P. Weisz, \Journal{\NPB}{240[FS12]}{349}{1984}.
\bibitem{vb}P. van Baal, \Journal{\PLB}{224}{397}{1989};
\Journal{\NPB}{351}{183}{1991}.
\bibitem{mic}C. Michael, G.A. Tickle and M.J. Teper, 
\Journal{\PLB}{207}{313}{1988}; C. Michael, \Journal{\NPB}{329}{225}{1990}.
\bibitem{var} M. Falcioni, e.a., \Journal{\PLB}{110}{295}{1982};
K. Ishikawa, G. Schierholz and M. Teper, \Journal{\PLB}{110}{399}{1982};
B. Berg, A. Billoire and C. Rebbi, \Journal{\ANP}{142}{185}{1982}.
\bibitem{kar}F. Karsch, \Journal{\NPB}{205[FS5]}{285}{1982}.
\bibitem{tep}M. Teper, \Journal{\PLB}{183}{345}{1987};
\Journal{\PLB}{185}{121}{1987}.
\bibitem{fs}A.M. Ferrenberg and R.H. Swendsen, \Journal{\PRL}{61}{2635}{1988}.
\end{thebibliography}
\end{document}